\documentstyle[12pt]{article}
\begin{document}
\bibliographystyle{unsrt}
\newcommand{\bra}[1]{\left < \halfthin #1 \right |\halfthin}
\newcommand{\ket}[1]{\left | \halfthin #1 \halfthin \right >}
\newcommand{\alp}{\mbox{\boldmath $\alpha$ \unboldmath}}
\newcommand{\sig}{\mbox{\boldmath $\sigma$ \unboldmath}}

\title{Longitudinal Transitions Of Baryon Resonances \\
In Constituent Quark Model}

\author{Zhenping Li$^1$, Yubing Dong$^2$ and Weihsing Ma$^2$ \\
\\
$^1$ Physics Department, Carnegie-Mellon University \\
Pittsburgh, PA. 15213-3890 \\
$^2$ CCAST(World Laboratory), P. O. Box 8730, Beijing 100080 \\
and\\
Institute of High Energy Physics, Academia Sinica\\
Beijing 100039, P.R. China}
\maketitle

\begin{abstract}
A longitudinal transition operator that satisfies the gauge
invariance requirement is introduced in constituent quark
 model.  The corresponding longitudinal transitions between
the nucleon and baryon resonances are calculated.  We show
that the study of the longitudinal coupling plays an important
role in understanding the structure of baryons.
\end{abstract}

\newpage
\subsection*{1. Introduction}
The electromagnetic transition between the nucleon and excited baryons
has been shown to be a very important probe to the structure of nucleon and
baryon resonances.  A significant
progress has been made since the theoretical investigations by Copley, Karl
and Obryk\cite{cko}, and Feynman, Kisslinger and Ravndal\cite{fkr}, who
presented first evidence of underlying $SU(6)\otimes O(3)$ symmetry of
baryon spectrum.  Recent studies have shown\cite{cl90} that the
relativistic effects are required in order
to give a consistent description of baryon spectrum and its transitional
properties, moreover, they are also essential to
 generate the model independent results in the low
energy limit, such as the low energy theorem in the Compton
scattering and corresponding Drell-Hearn-Gerasimov sum rule\cite{dhg}. The
 calculations including the relativistic effects in more realistic
potential quark models, such as
the Isgur-Karl\cite{ik1} model and its relativised version\cite{ci},
have shown that the successes of the nonrelativistic quark model
have been preserved, thus both spectroscopy and transitions of baryon
 resonances can be described in the same framework.

However, these studies have mostly concentrated on the transverse
transition amplitudes $A_{1/2}$ and $A_{3/2}$, and  there is an additional
longitudinal transition amplitude $S_{1/2}$ in the electroproduction
that has not been systematically studied.   Although attempts\cite{warn}
 have been made to investigate the longitudinal transition,
there is an important theoretical issue which was not treated
consistently in these investigations;
the usual expression
for the longitudinal transition operator in a quark system
\begin{equation}\label{1}
H^L_{em}=\epsilon_0 J_0-\epsilon_3 J_3,
\end{equation}
where
\begin{equation}\label{2}
J_0=\sum_j\sqrt {\frac {2\pi} {k_0}} e_j  e^{i{\bf k
\cdot r}_j}
\end{equation}
and
\begin{equation}\label{3}
J_3=\sum_j\sqrt {\frac {2\pi} {k_0}} e_j\frac 1{2m_q}
\left [ p_{3,j}e^{i{\bf k \cdot r}_j}
+e^{i{\bf k \cdot r}_j} p_{3,j}\right ]
\end{equation}
and quark $j$ at position ${\bf r}_j$ has mass and charge
$m_j$ and $e_j$,  does not satisfy the gauge--invariance constraint
\begin{equation}\label{4}
k_{\mu}J^{\mu}=0,
\end{equation}
where $k_{\mu}=\{k_0,\ 0,\ 0,\ k_3\}$ is the
momentum of the photon.  This was noticed some time
ago\cite{close}, and was emphasized by Bourdeau and Mukhopadhyay\cite{bm}
in their study of the transition $\gamma_v N\to \Delta$ in
the Isgur--Karl\cite{ik1} and the Vent--Baym--Jackson models\cite{ik}.
One solution\cite{close} to this problem is to add an
{\it ad hoc} current
\begin{equation}\label{5}
J^\prime_3 = - \frac {k_3 J_3-k_0 J_0}{ |k_3|^2} k_3
\end{equation}
to Eq. \ref{3}, which was used in the calculation of the
longitudinal transitions between the nucleon and baryon
resonances\cite{warn}.

The focus of this paper is to derive a longitudinal transition
operator that satisfies the gauge invariant condition, and use this
 transition operator to study the longitudinal transitions between
 the nucleon and baryon resonances.  In Ref. \cite{cl90}, we shown
that the current conservation in the nonrelativistic limit is
equivalent to the energy conservation with the nonrelativistic
kinematics.  Thus, in addition to the truncated model space problem
discussed in Ref. \cite{bm},  the current conservation will break down
due to the nonrelativistic treatment of the recoil effects.
In next section, we will show that the relativistic electromagnetic
current does satisfy the gauge invariant condition, assuming that
the wavefunction is  the eigenstate of the relativistic Hamiltonian.
Thus, the problem could be avoided  by imposing the current
conservation in the relativistic limit and then extracting the appropriate
nonrelativistic expression, the resulting longitudinal transition
operator will be gauge invariant.

In section 3, we calculate the longitudinal transition amplitudes
$S_{1/2}$ using the new transition operator derived in section 2.
We find significant differences between our results with those
 in Ref. \cite{warn},  in which the current in Eq. \ref{5} was added
to Eq. \ref{1}.  We present our results in terms of the ratio
between the longitudinal and the transverse cross sections, which
would be easier to compare with the experimental data.
Finally, the conclusions will be given in section 4.

\subsection*{2. A gauge invariant longitudinal transition operator}

The $H^L_{em}$ in Eq. \ref{1} follows from a
nonrelativistic approximation to the longitudinal quark--photon vertex,
\begin{equation}\label{6}
H^{rel}_{em}=\epsilon_0J^{rel}_0-\epsilon_3J^{rel}_3,
\end{equation}
where
\begin{equation}\label{7}
J_0^{rel}=\sqrt{\frac {2\pi}{k_0}} \sum_{j=1}^3
e_je^{i{\bf k\cdot r}_j}
\end{equation}
and
\begin{equation}\label{8}
J_3^{rel}= \sqrt{\frac {2\pi}{k_0}}
 \sum_{j=1}^3 e_j { \alpha_{3,j}}
  e^{i{\bf k\cdot r}_j},
\end{equation}
and $ \alpha_{3,j}$ are Dirac matrices. We can
rewrite the current $J_3$  as
\begin{eqnarray}\label{9}
J^{rel}_3 & = & \sqrt{\frac {4\pi}{2k_0} }
 \left [\sum_{j=1}^3 {\alp_j}\cdot {\bf p}_j
,\  \sum_{j=1}^3 e_j e^{i{\bf k\cdot r}_j}\frac 1{k_3}\right ]
 \nonumber \\ &= &\sqrt {\frac {4\pi}{2k_0} }
 \left [H_b,\  \sum_{j=1}^3 e_je^{i{\bf k\cdot  r}_j}\frac 1{ k_3}\right ],
\end{eqnarray}
where the Hamiltonian\cite{mn} in Eq. 2-11 for  a three-body system
is
\begin{eqnarray}\label{10}
H=\sum_{i=1}^3\left \{ \alp_i\cdot {\bf p}_i+\beta_im_i\right \}
+\sum_{i<j}\bigg \{ V_v({\bf r})
(1-\frac 12 \alp_i\cdot \alp_j)\nonumber \\ +
\frac 12 \alp_i\cdot {\bf r}\  \alp_j\cdot
{\bf r}\frac {V^\prime_v({\bf r})}{|{\bf r}|}
 +\beta_i\beta_j V_s({\bf r} \},
\end{eqnarray}
where ${\bf r}={\bf r}_i-{\bf r}_j$,
$V_v^\prime=\frac{dV_v({\bf r}) }{dr}$,
and $V_v({\bf r})$ and $V_s({\bf r})$ denote vector
 and scalar binding potentials for the
quark system.  Typically, $V_s({\bf r})$ could be a long
range scalar linear potential and $V_v({\bf r})$ a single-gluon exchange
potential.  Thus
\begin{eqnarray}\label{11}
\langle\Psi^{rel}_f
|J_3^{rel}|\Psi^{rel}_i\rangle  =\langle\Psi^{rel}_f
|\left [H_b,\  \sum_{j=1}^3 e_je^{i{\bf k\cdot r}_j}
\frac 1{k_3}\right ]
|\Psi^{rel}_i\rangle \nonumber \\
=(E^{rel}_f-E^{rel}_i)\langle\Psi^{rel}_f
|\sum_{j=1}^3 e_je^{i{\bf k\cdot r}_j}|\Psi^{rel}_i
\rangle \frac 1{k_3} \nonumber \\ =
\frac {(E^{rel}_f-E^{rel}_i)}{k_3}\langle\Psi^{rel}_f
|J_0^{rel}|\Psi^{rel}_i\rangle ,
\end{eqnarray}
where the initial- and final-state wavefunctions
$|\Psi^{rel}_i\rangle$ and  $|\Psi^{rel}_f\rangle$
must be eigenfunctions of the Hamiltonian $H_b$.
In a radiative transition the energy difference
between initial  and final states equals the photon
 energy, that is
\begin{equation}\label{13}
E^{rel}_f-E^{rel}_i=k_0.
\end{equation}
Note that Eq. \ref{13} is exact in relativistic limit,
so we have the gauge invariance constraint
\begin{equation}\label{14}
k^{\mu}J_{\mu}^{rel}=k_0J_0-k_3J_3=0.
\end{equation}
Since the currents $J^{rel}_0$ and $J_3^{rel}$ have different
transformation to the nonrelativistic limit, in particular the
nonrelativistic kinematics is used in Eq. \ref{13}, the
current are no longer conserved in the nonrelativistic limit.
Moreover, Eq. \ref{11} shows that the binding potential
plays an important role in the current $J^{rel}_3$,  thus the
truncated model space will further destroy the current conservation,
which has been discussed in detail in Ref. \cite{bm}.

This problem could be avoid if we take a different approach to
transform $H^{rel}_{em}$ in Eq. \ref{6} into the nonrelativistic
 limit;  since the current conservation is exact in
the relativistic limit, we substitute Eq. \ref{14} into Eq. \ref{6};
\begin{equation}\label{15}
H_{em}^{rel}=\left [\epsilon_0 -\epsilon_3 \frac {k_0}{k_3}
\right ]J_0^{rel},
\end{equation}
and we chose
the longitudinal polarization vector $\epsilon^L_{\mu}$
\begin{equation}\label{16}
\epsilon^L_{\mu}=\left \{ \epsilon_0,\ 0,\ 0, \epsilon_z\right \}
=\left \{ \frac {k_3}{\sqrt {Q^2}},\ 0,\ 0,\
\frac {k_0}{\sqrt {Q^2}}\right \},
\end{equation}
so the gauge invariant condition,
\begin{equation}\label{17}
k^{\mu}\cdot \epsilon_{\mu}=0,
\end{equation}
is satisfied ($Q$ is the virtual photon mass).
Note also that
\begin{equation}\label{18}
\epsilon_0 -\frac {\epsilon_zk_0}{k_z}
 =\frac {\sqrt {Q^2}}{k_z},
\end{equation}
which leads to
\begin{equation}\label{19}
\langle \Psi^{rel}_f| H^{rel}_{e,m}|\Psi^{rel}_i\rangle
=\sum_j\frac {\sqrt {Q^2}}{k_3} \langle \Psi^{rel}_f|J_0^{rel}
|\Psi^{rel}_i\rangle.
\end{equation}
Of course, the longitudinal electromagnetic interaction
should be proportional to $\sqrt {Q^2}$, and vanishes in the real
photon limit $Q^2=0$.  This is a direct consequence of
 the gauge invariance.

The nonrelativistic expansion of Eq. \ref{19} has been given in Ref.
\cite{cl93};
\begin{eqnarray}\label{20}
J_0 & = & \sqrt {\frac {2\pi} {k_0}}
\bigg \{\sum_j \left (e_j+
\frac {ie_j}{4m_j^2} {\bf k}\cdot (
{\sig_j}\times {\bf p}_j)\right ) e^{i{\bf k\cdot
 r}_j} \nonumber \\  & - & \sum_{j<l}\frac i{4M_T}\left ( \frac {
 \sig_j}{m_j}-\frac{
\sig_l}{m_l}\right)\cdot\left ( e_j {\bf k}\times
{\bf p}_le^{i{\bf k\cdot r}_j}-e_l {\bf k}\times {\bf p}_j e^{i{\bf k\cdot
r}_l}\right )\bigg \},
\end{eqnarray}
where the first term is the charge operator, which is
conventionally used in
the calculation of longitudinal helicity amplitudes;   the
second and third terms are spin--orbit and nonadditive terms
which have counterparts in the transverse electromagnetic
 transition\cite{cl90}.  The spin--orbit and nonadditive
 terms represent $O(v^2/c^2)$ relativistic
 corrections to the first term, which have long been known to
 be necessary even for systems of free particles, if low-energy
 theorem and Drell-Hearn-Gerasimov sum rule are to be
 satisfied\cite{cc} for the real photon case.
The longitudinal helicity amplitude $S_{\frac 12}$ is defined by
\begin{equation}\label{21}
S_{\frac 12}
 =\langle \Psi_f |J_0 |\Psi_i \rangle
\end{equation}
where $J_0$ is given by Eq. \ref{20}. The group structure of $J_0$ is
\begin{equation}\label{22}
J_0=A {\bf I} +B(S_+L_--S_-L_+),
\end{equation}
where ${\bf I}$ is the identity operator, $A$ and $B$ are the
coefficients determined by Eq. \ref{20}.  The second term
corresponds to the spin-orbit and nonadditive terms, and require
that the spin and orbital angular momentum z-component
 change by $\pm 1$ unit in a transition.
Thus, if there is no orbital angular momentum in the initial
and final state wavefunctions, the contribution from the second term
vanishes. In particular, the selection rule\cite{zl92} that the longitudinal
helicity amplitudes vanish for the transition between
the nucleon and hybrid states survives these relativistic
corrections if the quark spatial wavefunction of a hybrid
 state is essentially the same as that of the nucleon
and does not have orbital angular momentum.

It should be noted that the expression for $H_{em}$ may not be unique in
the nonrelativistic limit; for example, $H_{em}$ can also be written as
\begin{equation}\label{23}
H_{em}=\frac {\sqrt {Q^2}}{k_0} J_3
\end{equation}
due to the current conservation in the relativistic limit.
The nonrelativistic expression of $J_3$, however, is much more
complicated due to the explicit presence of the
binding potential shown in Eq. \ref{11},  and the problem of the
truncated model space becomes important.
Moreover, the recoil effects explicitly depend on the choice
 of the frame, which is also a source of the theoretical uncertainty.
This is why  Eq. \ref{21} is more convenient and
simpler to use, as the explicit dependence of the recoil effects on
the choice of frame is eliminated.

\subsection*{3. The Longitudinal Coupling of Baryon Resonances}

In Table 1, we show the analytical expressions of the longitudinal
transition between the nucleon and the baryon resonances in the
$SU(6)\otimes O(3)$ symmetry limit.
The terms proportional to $\frac {\alpha^2}{m_q^2}$ represent
the  relativistic contributions that come from the spin-orbit
and nonaddtive term in Eq. \ref{20}.  The relativistic effects
only scale the longitudinal coupling amplitudes, and do not
affect the general behaviour of $Q^2$ dependence of $S_{1/2}(Q^2)$.
   Therefore, the ratio
between the longitudinal couplings of the resonances $S_{11}(1530)$
and $D_{13}(1520)$ would be independent of $Q^2$
 since their masses are approximately equal.
This ratio is determined by the Clebsch-Gordon coefficients in the
nonrelativistic limit;
\begin{equation}\label{24}
\frac {S_{1/2}(S_{11}(1530))}{S_{1/2}(D_{13}(1520))}=-\frac 1{\sqrt{2}},
\end{equation}
and the relativistic effects would change
this ratio by a factor of $\frac {1+\frac {\alpha^2}{6m_q^2}}{1-
\frac {\alpha^2}{12m_q^2}}$.  The standard quark model parameters
$m_q=0.33$ GeV and $\alpha^2=0.17$ GeV$^2$ give
\begin{equation}\label{25}
\frac {1+\frac {\alpha^2}{6m_q^2}}{1-\frac {\alpha^2}{12m_q^2}}
=1.45 ,
\end{equation}
thus, this give an approximately $-1$ ratio with the relativistic
corrections.  The relativistic effects also lead to a nonzero
longitudinal transitions between the resonance $D_{15}(1670)$,
which vanishes for the nonrelativistic transition operator.
This gives us an important experimental test for the spin-orbit and
nonadditive term in the longitudinal transition operator.

The calculation of the $Q^2$ dependence of
$S_{1/2}(Q^2)$ follows the procedure of
Foster and Hughes\cite{fh}; a Lorentz boost factor in the spatial
integrals are introduced so that
\begin{equation}\label{26}
R(k) \to \frac 1{\gamma^2} R\left ( \frac k {\gamma}\right ),
\end{equation}
where the Lorentz boost factor $\gamma$ can be written as
\begin{equation}\label{27}
\gamma^2=1+\frac {k^2}{(M_r+M_p)^2}
\end{equation}
in the equal velocity frame and
\begin{equation}\label{29}
k^2(EVF)=\frac {(M_r^2-M_p^2)^2}{2M_rM_p}+
\frac {Q^2(M_r^2+M_p^2)}{4M_rM_p}
\end{equation}
for the initial nucleon mass $M_p$ and final resonance mass $M_r$.
The corresponding $Q^2$ dependence of $S_{1/2}(Q^2)$ for $S_{11}(1530)$
is given by
\begin{equation}\label{28}
S_{1/2}(Q^2)=\frac 23\sqrt{\frac {\pi}{k_0}}\mu m_q \frac {k}{\gamma^3\alpha}
\left (1+\frac {\alpha^2}{6m_q^2}\right )\frac 1{1+Q^2/0.8}
e^{-\frac {k^2}{6\alpha^2\gamma^2}},
\end{equation}
where an {\it ad hoc} form factor $\frac 1{1+Q^2/0.8}$ is being added,
since it gives a better quantitative description of the $Q^2$ dependence
of transverse helicity amplitudes\cite{fh} for $\alpha^2=0.17$ GeV$^2$,
and it becomes unnecessary with $\alpha^2=0.09$ GeV$^2$\cite{cl90}.
In Fig.1, we show the $Q^2$ dependence of the longitudinal
amplitude $S_{1/2}(Q^2)$ for the resonance $S_{11}(1530)$ in the
$SU(6)\otimes O(3)$ symmetry limit.  The relativistic effects
increase $S_{1/2}(Q^2)$ by about 25 percent. The resulting $S_{1/2}(Q^2)$
is significantly larger than that in Ref. \cite{warn}, and in better
 agreement with the analysis by Gerhardt\cite{ger}, who extracted
the longitudinal transition amplitudes from the electroproduction data.
This shows the importance of choosing the correct transition
operator for the longitudinal coupling.

A more important  quantity is the ratio between the longitudinal
coupling and transverse coupling amplitudes,
\begin{equation}\label{30}
R=\frac {S^2_{1/2}(Q^2)}{A^2_{1/2}(Q^2)+A^2_{3/2}(Q^2)},
\end{equation}
in which the common factors, such as the {\it ad hoc} form factor in
Eq. \ref{28}, may cancel out, thus provides us a direct probe of the
underlying structure of the resonance.   The analytic expressions for
transverse helicity amplitudes $A_{1/2}$ and $A_{3/2}$ are given in
Ref. \cite{cl90}.   The $Q^2$ dependence of this
ratio for the resonance $S_{11}(1530)$ is shown in Fig. 2, it shows
a strong presence of the longitudinal transitions for this resonance.
Moreover, the result for the transverse
helicity amplitude $A_{1/2}(Q^2)$\cite{cl90} in the symmetry limit
is twice larger than the experimental data and it decreases too fast as
$Q^2$ increases, this indicates a strong configuration mixing
for the resonance $S_{11}(1530)$.  The data of the $Q^2$ dependence
of this ratio provide another crucial
test to the various potential quark models with  different
binding potentials.  The extension of this investigation to include
the configuration mixing is in progress.  It is interesting to note
that the configuration mixings in the Isgur-Karl model\cite{cl90} do not
change the result of naive quark model significantly due to the strong
presence of the $70$ multiplet state in the nucleon wavefunction.

It is straightforward to obtain the $Q^2$ dependence of the
longitudinal transition for the resonance $D_{13}(1520)$,
which Eq. \ref{25} shows that its longitudinal transition
approximately equals to that of the resonance $S_{11}(1520)$
with a opposite sign.   Thus, strong contributions from the
longitudinal transitions are expected for the P-wave resonances,
and furthermore, the relativistic effects contributes significantly
to $S_{1/2}(Q^2)$ because of the nonzero orbital angular momentum
in the wavefunction of P-wave resonances.

In Fig. 3, we show the $Q^2$ dependence of $S_{1/2}(Q^2)$
for the resonance $F_{16}(1688)$,  the relativistic effects
reduces the longitudinal transitions significantly, which are
in better agreement with the result of Gerhardt\cite{ger}.
The $Q^2$ dependence of the longitudinal and transverse transitions
for this resonance is shown in Fig. 4; the longitudinal
transitions are much smaller in this case, in particular,
it is less than 0.1 with the relativistic corrections.
The naive quark model\cite{cko,cf} does give a good description
of the transverse helicity amplitudes even quantitatively.
More precise data for the longitudinal transition would provide
us more insights into the structure of this resonance;  deviations
from the prediction in Figs. 3 and 4 would be evidences for
the configuration mixings.

The longitudinal transition between the resonance $P_{33}(1232)$ and
the nucleon  vanishes in the symmetry limit.
However, if there is
a small component of the orbital angular momentum in both wavefunctions
of the nucleon and the resonance $P_{33}(1232)$, the spin-orbit and the
nonadditive term in the longitudinal transition operator would
lead to a nonzero longitudinal transition between these two states.
In Fig. 5, we show the ratio
\begin{equation}\label{31}
R=-\frac {S_{1/2}(Q^2)}{\sqrt{2} M1}=\frac {\sqrt{2}S_{1/2}(Q^2)}
{\sqrt {3}A_{3/2}(Q^2)+A_{1/2}(Q^2)}
\end{equation}
(notice $S_{1/2}(Q^2)$ in Eq. \ref{21} differs by a factor of
$-\frac 1{\sqrt{2}}$ from Ref. \cite{bm}) for the Isgur-Karl
model, whose wavefunctions
for the nucleon and the resonance $P_{33}(1232)$ are obtained from
Ref. \cite{ikk}, and the transverse helicity amplitudes $A_{3/2}(Q^2)$
and $A_{1/2}(Q^2)$ are given in Ref. \cite{cl90}.
We find that the relativistic corrections approximately double this
ratio.  While the experimental data\cite{volker} are inconclusive
for a finite ratio $E1/M1$, they do suggest a finite and negative ratio
$S_{1/2}/M1$. This suggests that the longitudinal transitions
between the nucleon and the resonance $P_{33}(1232)$ might be a better
probe to the orbital angular momentum in the nucleon wavefunction
than the $E1$ transition, which certainly deserves  more attention.

\subsection*{4. Conclusions}

We have derived a gauge independent longitudinal transitions
operator, in which the relativistic effects are included.
The calculations of longitudinal transitions between the
nucleon and baryon resonances are made in the symmetry limit.
We show that the relativistic effects in the transition
operator give important contributions to the
longitudinal helicity amplitudes, especially in the
transition between the nucleon and the resonance $P_{33}(1232)$.

We show that the longitudinal transitions of baryon resonances
play an important role in understanding the structure
of baryons.  The longitudinal transition amplitude $S_{1/2}(Q^2)$
decreases  as the $Q^2$ increases.
Thus, it is important in the small $Q^2$
regions, in particular, for the transitions between the
nucleon and the P-wave resonance, which is accesible to the
experiments at CEBAF.

An extension of this investigation is to study the
spin structure function $g_{1,2}(x,Q^2)$ in the resonance
region,  where the studies\cite{ll94} have shown that the
$Q^2$ dependence of the spin-structure function is very
significant, and the longitudinal transitions amplitudes
provide important contributions to the spin-structure
functions in the low $Q^2$ region.

This work was supported in part
by the United States National Science Foundation grant PHY-9023586.

\newpage
\begin{table} {Table 1: Transition matrix elements between the nucleon and
baryon resonances in the $SU(6)\otimes O(3)$ symmetry limit.  The full
matrix elements are obtained by multiplying the entries in this table by a
factor $\sqrt{\frac {2\pi}{k_0}} 2\mu m_q e^{-\frac {{\bf k}^2}{6\alpha^2}}$,
and $S_{\frac 12}^n=S_{\frac 12}^p$ for $\Delta$ states.}
\\ [1ex]
\begin{tabular}{rrcll}\hline \hline
Multiplet & States &  & Proton & Neutron  \\ \hline
$[70,1^{-}]_1$ & $N(^2P_M){\frac 12}^{-1}$ & & $\frac 1{3\sqrt {2}}\frac
{|\bf k|}{\alpha}\left (1+\frac {\alpha^2}{6m_q^2}\right )$ & $-\frac 1
{3\sqrt {2}}\frac {|\bf k|}{\alpha} \left (1+\frac {\alpha^2}{6m_q^2}\right )$
\\ [1ex]
& $N(^2P_M){\frac 32}^{-1}$ & & $-\frac 13 \frac {|\bf k|}{\alpha}\left
 (1-\frac {\alpha^2}{12m^2_q}\right )$ & $\frac 13\frac {|\bf k|}{\alpha}
\left (1-\frac {\alpha^2}{12m^2_q}\right )$ \\ [1ex]
& $N(^4P_M){\frac 12}^{-1}$ & & $\frac 1{36\sqrt{2}} \frac {\alpha |{\bf k}|}
{m^2_q}$ & $-\frac 1{108\sqrt{2}} \frac {\alpha |{\bf k}|}
{m^2_q}$ \\ [1ex]
& $N(^4P_M){\frac 32}^{-1}$ & & $\frac 1{9\sqrt{10}} \frac {\alpha |{\bf k}|}
{m^2_q}$ & $-\frac {5}{27\sqrt{10}} \frac {\alpha |{\bf k}|}
{m^2_q}$ \\ [1ex]
& $N(^4P_M){\frac 52}^{-1}$ & & $\frac 1{12\sqrt{10}} \frac {\alpha |{\bf k}|}
{m^2_q}$ & $-\frac {5}{36\sqrt{10}} \frac {\alpha |{\bf k}|}
{m^2_q}$ \\ [1ex]
& $\Delta(^2P_M){\frac 12}^{-1}$ & & $-\frac 1{3\sqrt{2}} \frac {|{\bf k}|}
{\alpha}\left (1-\frac {\alpha^2}{6m^2_q}\right )$ &
 \\ [1ex]
& $\Delta(^2P_M){\frac 32}^{-1}$ & & $\frac 1{3} \frac {|{\bf k}|}
{\alpha}\left (1+\frac {\alpha^2}{12m^2_q}\right )$ &
\\ [1ex]
$[56,0^+]_2$ & $N(^2S_{S^\prime}){\frac 12}^+$ & &
$-\frac 1{3\sqrt{6}}\frac {{\bf k}^2}{\alpha^2}$ & 0 \\ [1ex]
& $\Delta(^4S_{S^\prime}){\frac 32}^+$ & & 0 & \\ [1ex]
$[56,2^+]_2$& $N(^2D_{S}){\frac 32}^+$ & & $-\frac 1{3\sqrt{15}}\frac {{\bf
k}^2}{\alpha^2}
\left (1+\frac {\alpha^2}{2m_q^2}\right )$ & $-\frac {{\bf k}^2}
{12\sqrt{15}m_q^2}$ \\ [1ex]
& $N(^2D_{S}){\frac 52}^+$ & & $-\frac 1{3\sqrt{10}}\frac {{\bf k}^2}{\alpha^2}
\left (1-\frac {\alpha^2}{3m_q^2}\right )$ & $\frac {{\bf k}^2}
{9\sqrt{10}m_q^2}$ \\ [1ex]
& $\Delta(^4D_S){\frac 12}^+$ &  & $-\frac {5{\bf k}^2}{72\sqrt{15}m_q^2}$ &
\\ [1ex]
& $\Delta(^4D_S){\frac 32}^+$ & & 0 & \\ [1ex]
& $\Delta(^4D_S){\frac 52}^+$ & & $\frac {5\sqrt{5}{\bf k}^2}
{216\sqrt{7}m_q^2}$ & \\ [1ex]
& $\Delta(^4D_S){\frac 72}^+$ & & $\frac {5{\bf k}^2}{36\sqrt{105}m_q^2}$ &
\\ [1ex]
$[70,0^+]_2$ & $N(^2S_{M^\prime}){\frac 12}^+$ & &
$\frac 1{18}\frac {{\bf k}^2}{\alpha^2}$ &
$-\frac 1{18}\frac {{\bf k}^2}{\alpha^2}$ \\ [1ex]
\hline
\end{tabular}
\end{table}

\newpage
\section*{Figure Caption}
\begin{itemize}
\begin{enumerate}
\item The $Q^2$ dependence of $S_{1/2}(Q^2)$ for the resonance
$S_{11}(1535)$, the solid line represents the nonrelativistic result
and the dashed line includes the relativisitc corrections.
\item The ratio between the logitudinal and transverse cross sections
for the resonance $S_{11}(1535)$.
The solid line represents the nonrelativistic result, and the dashed
line includes the relativistic corrections.
\item The same as Fig. 1 for the resonance
$F_{15}(1688)$.
\item The same as Fig. 2
for the resonance $F_{15}(1688)$.
\item The ratio between the longitudinal and $M1$ transitions for the
resonance $P_{33}(1232)$ in the Isgur-Karl model, see text.
\end{enumerate}
\end{itemize}

\begin{thebibliography}{99}
\bibitem{cko} L.A. Coplay, G. Karl and E. Obryk, Nucl. Phys. {\bf B13},
303(1969).
\bibitem{fkr} R. Feynamn, M. Kisslinger and F. Ravndal, Phys. Rev.
{\bf D3}, 2706(1971).
\bibitem{cl90} F.E.Close and Zhenping Li,   Phys. Rev.
 {\bf D42}, 2194 (1990), {\it ibid} {\bf D42}, 2207 (1990).
\bibitem{dhg} S. D. Drell and A. C. Hearn, Phys. Rev. Lett. {\bf 16},
908(1966); S. B. Gerasimov, Yad. Fiz. {\bf 2}, 839(1965) [Sov. J. Nucl. Phys.
{\bf 2}, 598(1966)].
\bibitem{ik1} N. Isgur and G. Karl, Phys. Rev. {\bf D18}, 4187(1978),
{\bf D19}, 2194 (1979).
\bibitem{ci} S. Capstick and N. Isgur, Phys. Rev. {\bf D34}, 2704(1986).
\bibitem{warn} M.Warns et al., {\it Z. Phys.} {\bf C45}, 613(1989);
 {\bf C45}, 627 (1989).
\bibitem{bm} M. Bourdeau and N. Mukhopadhyay, Phys. Rev. Lett. {\bf 58},
976(1987).
\bibitem{ik}V. Vento, G. Baym, and A.D. Jackson, Phys. Lett. {\bf 102B},
97 (1981);
\newline V. Vento and J. Navarro, Phys. Lett. {\bf 141B}, 28 (1984).
\bibitem{close} T. Abdullah and F. E. Close, Phys. Rev. {\bf D5}, 2332 (1972).
\bibitem{mn} R. McClary, and N. Byers, Phys. Rev. {\bf D28}, 1692 (1983).
\bibitem{cl93} F. E. Close and Zhenping Li, Phys. Lett. {\bf B289}, 143(1992).
\bibitem{cc} Zhenping Li, Phys. Rev. {\bf D47}, 1854(1993);
F.E. Close and L.A. Copley, Nucl. Phys. {\bf B19}, 477(1970);
F.E. Close and H. Osborn,  Phys. Rev. {\bf D2}, 2127 (1970).
\bibitem{zl92} Zhenping Li, V. Burkert and Zhujun Li, Phys. Rev. {\bf D46},
70(1992).
\bibitem{fh} F. Foster and G. Hughes, Z. Phys. {\bf C41}, 123(1982).
\bibitem{ger} C. Gerhardt, Z. Phys. {\bf C4}, 311(1980).
\bibitem{cf} F. E. Close and F. J. Gillman, Phys. Lett. {\bf B38}, 514(1972).
\bibitem{ikk} N. Isgur, G. Karl and R. Koniuk, Phys. Rev. Lett. {\bf 41},
1269(1978).
\bibitem{volker} N. C. Mukhopadhyay, {\it Excited Baryon 1988}, in
proceedings of the Topical Workshop, Troy, New York, edited by
G. Adams, N. C. Mokhopadhyay, and Paul Stoler (World Scientific, Singapore
1989); O. A. Rondon-Aramayo, Nucl. Phys. A490, 643(1988).
\bibitem{ll94} Zhenping Li and Zhujun Li, to be published in Phys. Rev.
D.
\end{thebibliography}
\end{document}